\documentstyle[11pt,epsf]{article}

\def\indspace{\hspace*{1.0em} }
\def\appendix{\setcounter{section}{0}
\def\thesection{Appendix \Alph{section}}
\def\theequation{\Alph{section}.\arabic{equation}}}

\newfont{\subsub}{cmr6}

\newcounter{szk}

\begin{document}
\title{Dynamical change of Pareto index in Japanese land prices
}
\author{
\footnote{e-mail address: ishikawa@kanazawa-gu.ac.jp} Atushi Ishikawa
\\
Kanazawa Gakuin University, Kanazawa 920-1392, Japan
}
\date{}
\maketitle

\begin{abstract}
\indent
We investigate the dynamical behavior in the large scale region
of non-equilibrium systems,
by employing data on the assessed value of land
in 1983 -- 2006 Japan.
In the system
we find the detailed quasi-balance,
which has the symmetry: 
$x_1 \leftrightarrow a~ {x_2}^{\theta}$
($x_1$ and $x_2$ are two successive land prices).
By using the detailed quasi-balance and Gibrat's law,
we derive Pareto's law
with varying Pareto index annually.
The parameter $\theta$ corresponds with the ratio of Pareto indices 
$(\mu_1 + 1)/(\mu_2 + 1)$,
and 
the relation is confirmed in the empirical data nicely.
\end{abstract}
\begin{flushleft}
\end{flushleft}

\vspace{1cm}
\section{Introduction}
\label{sec-Introduction}
\indspace

In the large scale region of wealth, income, 
profits, assets, sales, the number of employees and etc ($x$),
the cumulative probability distribution $P(> x)$ obeys a power-law:
\begin{eqnarray}
    P(> x) \propto x^{-\mu}~.
    \label{Pareto}
\end{eqnarray}
This power-law and the exponent $\mu$ are called 
Pareto's law and Pareto index, respectively~\cite{Pareto}.

Recently,
Fujiwara et al.~\cite{FSAKA} have explained Pareto's law (and the reflection law)
by using the law of detailed balance and Gibrat's law~\cite{Gibrat},
which are also observed in empirical data.
The detailed balance is time-reversal symmetry observed in a relatively stable economy:
\begin{eqnarray}
    P_{1 2}(x_1, x_2) = P_{1 2}(x_2, x_1)~.
    \label{Detailed balance}
\end{eqnarray}
Here $x_1$ and $x_2$ are  two successive incomes, 
profits, assets, sales, etc. and
$P_{1 2}(x_1, x_2)$ is a joint probability distribution function (pdf).
On the other hand,
Gibrat's law is valid in the large scale region where
the conditional probability distribution of growth rate $Q(R|x_1)$ 
is independent of the initial value $x_1$: 
\begin{eqnarray}
    Q(R|x_1) = Q(R)~.
    \label{Gibrat}
\end{eqnarray}
Here growth rate $R$ is defined as the ratio $R = x_2/x_1$ and
$Q(R|x_1)$ is defined by using the pdf $P_1(x_1)$
and the joint pdf $P_{1 R}(x_1, R)$ as $Q(R|x_1) = P_{1 R}(x_1, R)/P_1(x_1)$~.
In the proof, Fujiwara et al. assume no model and only use
these two underlying laws in empirical data.
In Ref.~\cite{Ishikawa2}, it is reported that
the Pareto index is also induced 
from the reflection law.

These findings are important
for the progress of econophysics.
Above derivations are, however, valid only in the economic equilibrium
where the detailed balance (\ref{Detailed balance}) holds.
In order to discuss the transition,
the dynamics should be established
by investigating long-term economic data in which dynamical transitions are observed. 
Unfortunately, it is difficult to obtain personal income or company size data
for a long period.

In this study,
we investigate the dynamical behavior in the large scale region
of non-equilibrium systems,
by employing data on the assessed value of land
in 1983 -- 2006 Japan.
Because
the distribution of Japanese land prices
has similar features with one of personal income and company size~\cite{Kaizoji},
and the long-term database is readily available~\cite{Web}.

In the non-equilibrium system
we find the detailed quasi-balance,
which has the symmetry: 
$x_1 \leftrightarrow a~ {x_2}^{\theta}$.
By using the detailed quasi-balance and Gibrat's law,
we derive Pareto's law
with varying Pareto index annually.
The parameter $\theta$ corresponds with the ratio of Pareto indices 
$(\mu_1 + 1)/(\mu_2 + 1)$,
and 
the relation is confirmed in the empirical data nicely~\cite{Ishikawa4}.
\section{Detailed quasi-balance}
\label{sec-Detailed quasi-balance}
\indspace
In Japan, land is a very important asset and
land prices change annually in a 24-period (1983 -- 2006).
This period contains bubble term (1986 -- 1991)
caused by the abnormal rise of land prices.
The economy correlates with land prices.
We employ the database of the assessed value of land,
which indicates the standard land prices,
covering the 24-year period from 1983 to 2006.\footnote{
In Ref.~\cite{Ishikawa4}, 
the number of data points of land prices increased gradually,
because the database only contained data points
which existed in the 2005 evaluation.
In this study, the database contains all data points
which existed in every year evaluation.
The results, however, do not change seriously.
}

\begin{figure}[hbp]
 \centerline{\epsfxsize=0.8\textwidth\epsfbox{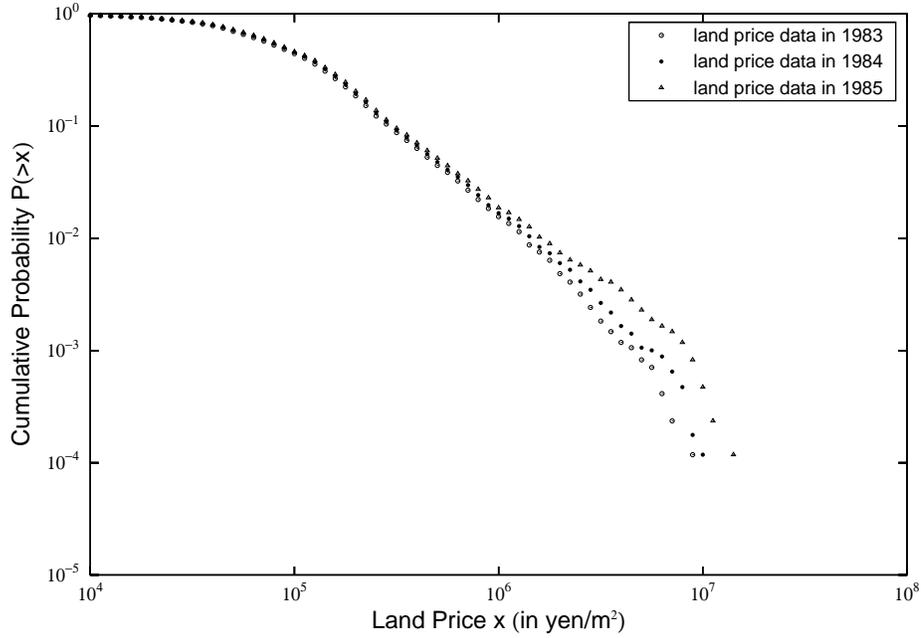}}
 \caption{Cumulative probability distribution $P(>x)$ of land prices
 in 1983 -- 1985.
 The number of the data points is ``16,975'' in all cases.
 Data points are equally spaced in logarithm of land price.}
 \label{PreBubbleDistribution}
\end{figure}

\begin{figure}[hbp]
 \centerline{\epsfxsize=0.8\textwidth\epsfbox{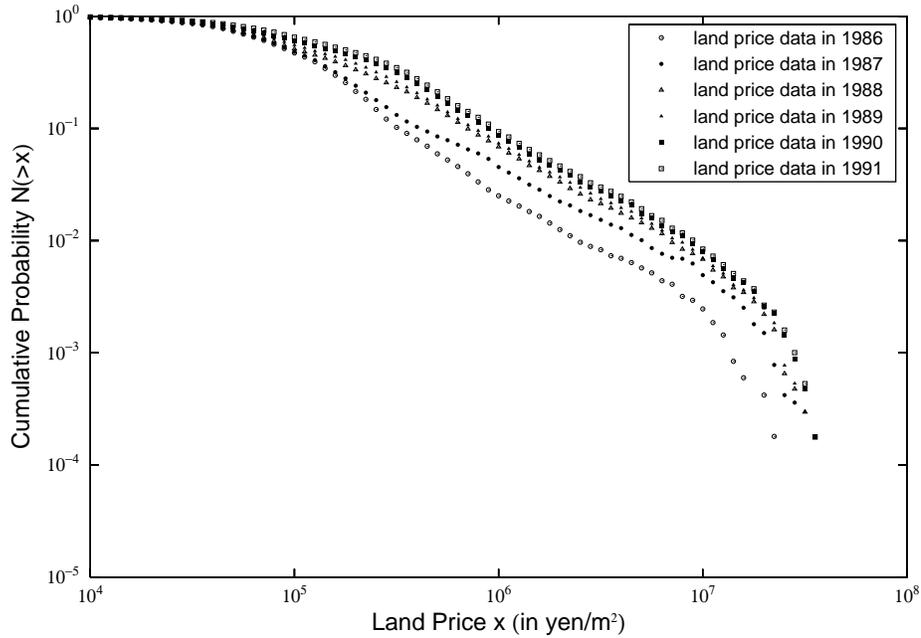}}
 \caption{Cumulative probability distribution $P(>x)$ of land prices
 in 1986 -- 1991.
 The number of the data points is ``16,635``, ``16,635'', ``16,820'',
``16,840'', ``16,865'' and ``16,892'', respectively.
Data points are equally spaced in logarithm of land price.}
 \label{BubbleDistribution}
\end{figure}

\begin{figure}[hbp]
 \centerline{\epsfxsize=0.8\textwidth\epsfbox{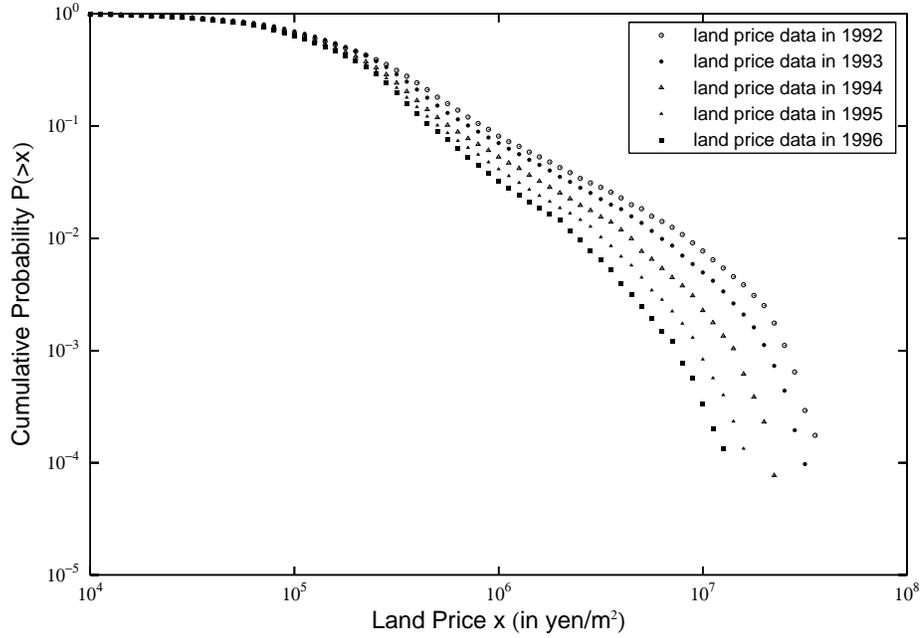}}
 \caption{Cumulative probability distribution $P(>x)$ of land prices
 in 1992 -- 1996.
 The number of the data points is ``17,115'', ``20,555'',
``26,000'', ``30,000'' and ``30,000'', respectively.
Data points are equally spaced in logarithm of land price.}
 \label{AfterBubbleDistribution}
\end{figure}

\begin{figure}[hbp]
 \centerline{\epsfxsize=0.8\textwidth\epsfbox{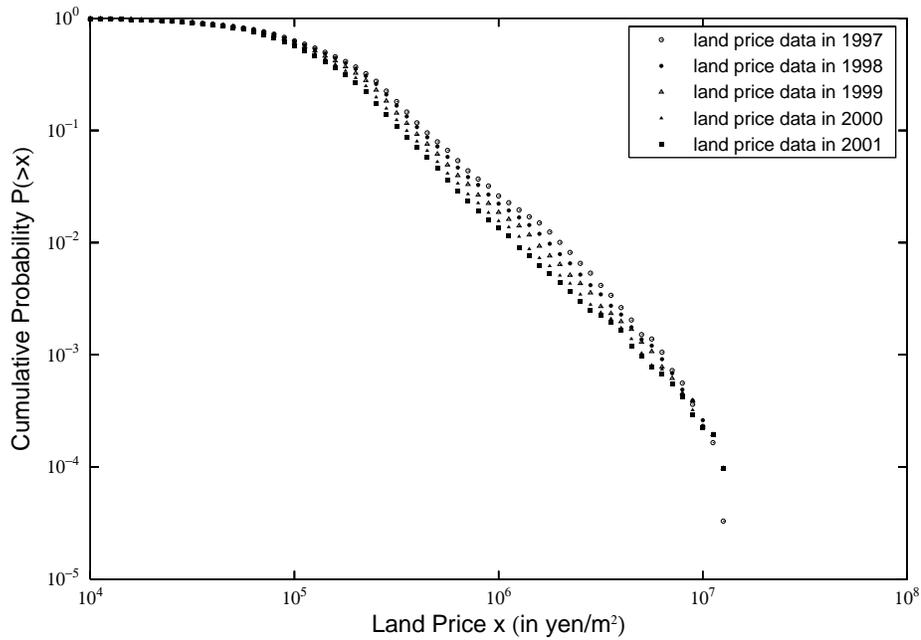}}
 \caption{Cumulative probability distribution $P(>x)$ of land prices
 in 1997 -- 2001.
 The number of the data points is ``30,300'', ``30,600'',
``30,800'', ``31,000'' and ``31,000'',  respectively.
Data points are equally spaced in logarithm of land price.}
 \label{StaticDistribution}
\end{figure}

\begin{figure}[hbp]
 \centerline{\epsfxsize=0.8\textwidth\epsfbox{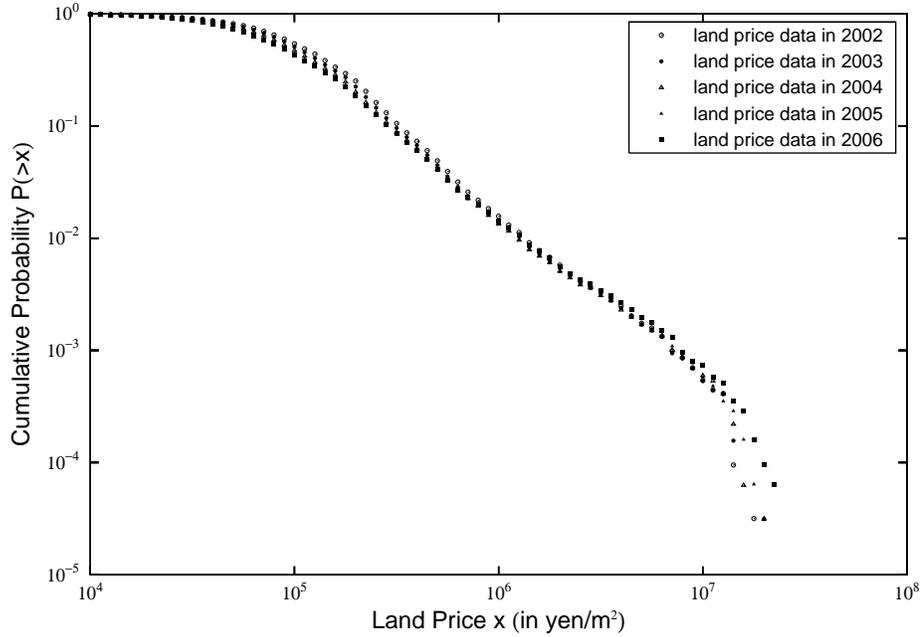}}
 \caption{Cumulative probability distribution $P(>x)$ of land prices
 in 2002 -- 2006.
 The number of the data points is  ``31,520'', ``31,866'',
``31,866'', ``31,230'' and ``31,230'', respectively.
Data points are equally spaced in logarithm of land price.}
 \label{quasiStaticDistribution}
\end{figure}

\begin{figure}[hbp]
 \centerline{\epsfxsize=0.8\textwidth\epsfbox{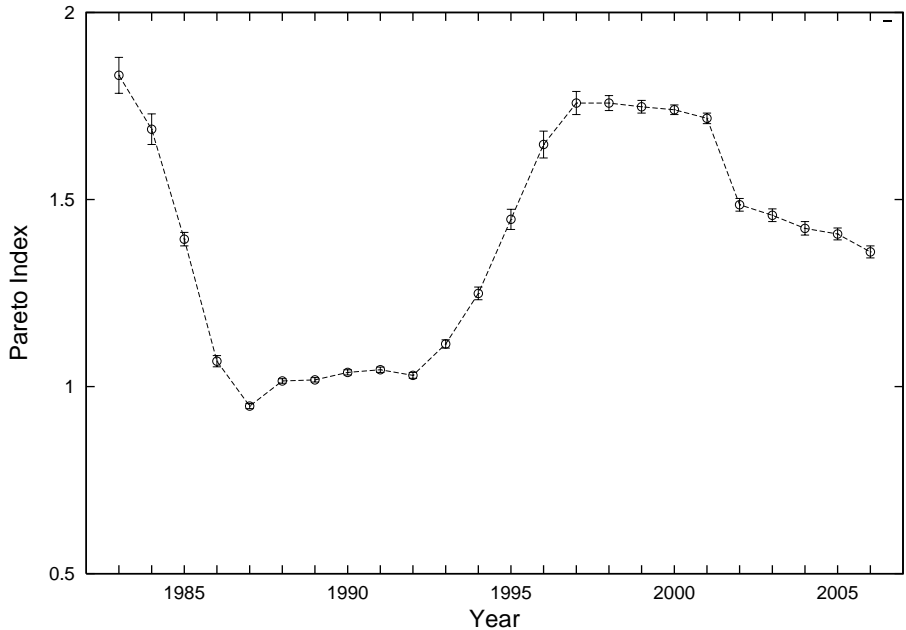}}
 \caption{Annual change of Pareto index $\mu$  from 1983 to 2006.
 For each year, Pareto index $\mu$ is estimated in the range of land prices from 
 $2 \times 10^5$ to $10^7~{\rm yen}/m^2$ by using least-squares fit to data points
 equally spaced in logarithm of land price
 in cumulative probability distributions 
 (Fig.~\ref{PreBubbleDistribution} -- \ref{quasiStaticDistribution}).
 }
 \label{VaryingParetoIndex}
\end{figure}

\begin{figure}[htb]
 \centerline{\epsfxsize=0.8\textwidth\epsfbox{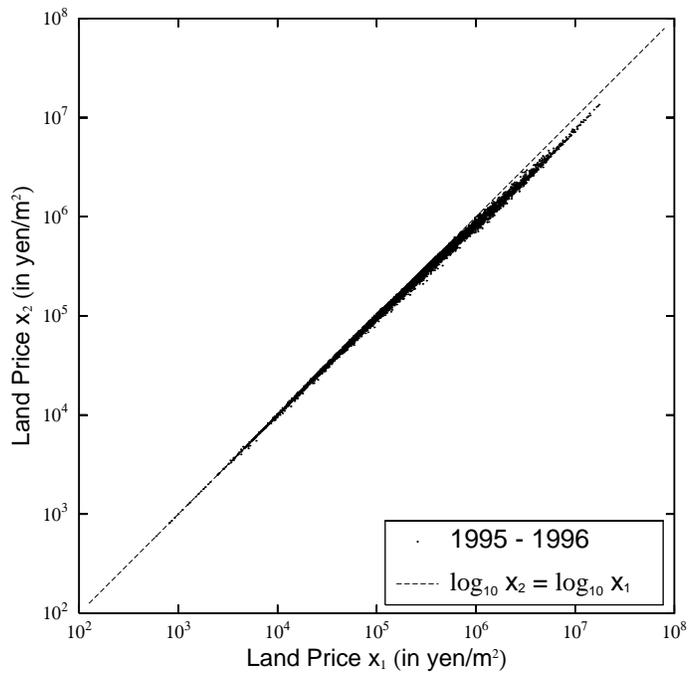}}
 \caption{The scatter plot of all pieces of land assessed in 1995 ($x_1$)
 and 1996 ($x_2$).
 The number of data points
 is ``11,278''.}
 \label{95vs96}
\end{figure}

\begin{figure}[htb]
 \centerline{\epsfxsize=0.8\textwidth\epsfbox{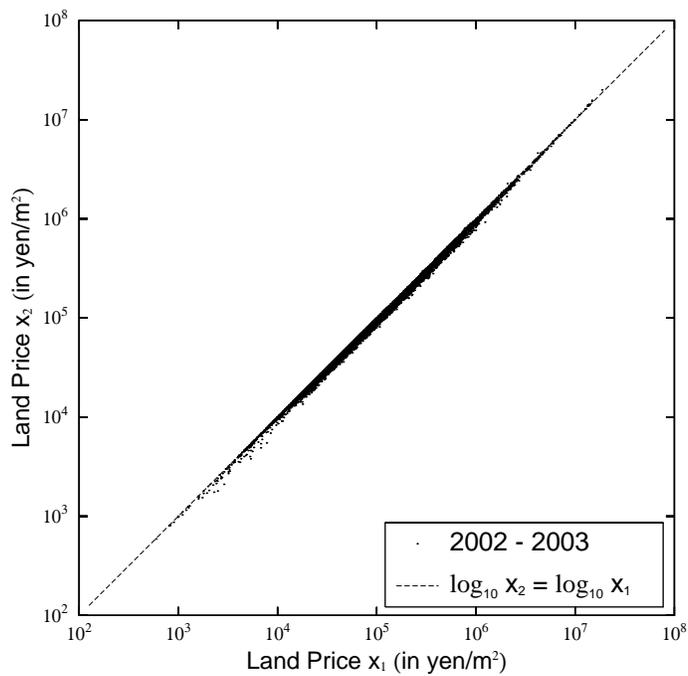}}
 \caption{The scatter plot of all pieces of land assessed in 2002 ($x_1$)
 and 2003 ($x_2$).
 The number of data points
 is ``6,839''.}
 \label{02vs03}
\end{figure}

The cumulative probability distributions of land prices are shown in
Fig.~\ref{PreBubbleDistribution} -- \ref{quasiStaticDistribution}.
From Fig.~\ref{PreBubbleDistribution} -- \ref{quasiStaticDistribution},
the power-law is confirmed in the large scale region.
For each year, we estimate Pareto index $\mu$ in the range of land prices from
$2 \times 10^5$ to $10^7~{\rm yen}/m^2$
where Pareto's law holds approximately.
Annual change of Pareto index $\mu$ from 1983 to 2006 is represented
in Fig.~\ref{VaryingParetoIndex}.
In this period, Pareto index has changed annually.
This means that the system is not in equilibrium
and the detailed balance (\ref{Detailed balance}) does not hold.
Actually, the breakdown in the large scale region
is observed in the scatter plot
of all pieces of land assessed in the database
(Fig.~\ref{95vs96} for instance).
There is no $x_1 \leftrightarrow x_2$ symmetry in 
Fig.~\ref{95vs96} obviously.
On the other hand,
the detailed balance ($x_1 \leftrightarrow x_2$ symmetry)  in the large scale region
is observed approximately in Fig.~\ref{02vs03}
for instance.

\begin{figure}[htb]
 \centerline{\epsfxsize=0.8\textwidth\epsfbox{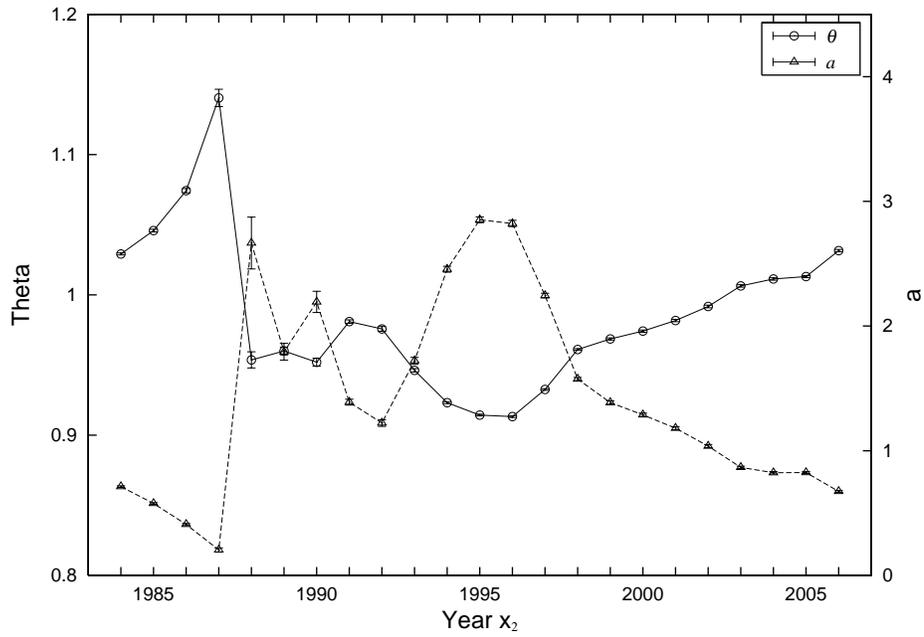}}
 \caption{Annual change of $\theta$ and $a$ of Eq.~(\ref{DqB}) in
 the year $(x_1, x_2) = (1983, 1984)$ -- $(2005, 2006)$.
 Because we demand the detailed quasi-balance (\ref{Detailed quasi-balance}) in the system,
 the two parameters $\theta$ and $a$ change in opposite direction.
 }
 \label{DqB}
\end{figure}

From Fig.~\ref{95vs96} -- \ref{02vs03},
we make a simple assumption that
the symmetry of the joint pdf $P_{1 2}(x_1, x_2)$ is represented
as a regression line fitted by least-square method as follows
\begin{eqnarray}
    \log_{10} x_2 = \theta~\log_{10} x_1 + \log_{10} a~.
    \label{Line}
\end{eqnarray}
In this form,
the detailed balance (\ref{Detailed balance}) has the special symmetry, 
$\theta = a = 1$.
For each scatter plot, we measure $\theta$, $a$
in the same range where Pareto index $\mu$ is estimated
and the result is shown in Fig.~\ref{DqB}.

From this symmetry ($a {x_1}^{\theta} \leftrightarrow x_2$), 
we extend the detailed balance (\ref{Detailed balance}) to
\begin{eqnarray}
    P_{1 2}(x_1, x_2) 
    = P_{1 2}( \left( \frac{x_2}{a} \right)^{1/{\theta}}, a~{x_1}^{\theta})~.
    \label{Detailed quasi-balance}
\end{eqnarray}
We call this law the detailed quasi-balance.

\section{Pareto's law with varying Pareto index}
\label{sec-Pareto's law with varying Pareto index}
\indspace
In this section,
we derive 
Pareto's law with varying Pareto index
by using the detailed quasi-balance (\ref{Detailed quasi-balance}).
In the proof, we assume Gibrat's law (\ref{Gibrat})
in the large scale region,
because the number of data points is insufficient to observe Gibrat's law.

Due to the relation of
$P_{1 2}(x_1, x_2)dx_1 dx_2 = P_{1 R}(x_1, R)dx_1 dR$
under the change of variables from $(x_1, x_2)$ to $(x_1, R)$,
these two joint pdfs are related to each other
\begin{eqnarray}
    P_{1 R}(x_1, R) = {x_1}^{\theta} P_{1 2}(x_1, x_2)~,
\end{eqnarray}
where we use a modified ratio $R \equiv x_2/{x_1}^{\theta}$.
From this relation, the detailed quasi-balance (\ref{Detailed quasi-balance})
is rewritten in terms of $P_{1 R}(x_1, R)$ as follows:
\begin{eqnarray}
    P_{1 R}(x_1, R) 
    = a R^{-1} P_{1 R}(\left( \frac{x_2}{a} \right)^{1/{\theta}}, a^2 R^{-1})~.
\end{eqnarray}
Substituting the joint pdf $P_{1 R}(x_1, R)$ for the conditional probability $Q(R|x_1)$,
the detailed quasi-balance is expressed as
\begin{eqnarray}
    \frac{P_1(x_1)}{P_1(\left( {x_2}/{a} \right)^{1/{\theta}})} 
    &=& \frac{a}{R} 
    \frac{Q(a^2 R^{-1}|\left( {x_2}/{a} \right)^{1/{\theta}})}{Q(R|x_1)}
    \label{DqB and Gibrat0}\\
    &=& \frac{a}{R} 
    \frac{Q(a^2 R^{-1})}{Q(R)} \equiv G(a)~.
    \label{DqB and Gibrat}
\end{eqnarray}

By expanding Eq.~(\ref{DqB and Gibrat}) around $R=a$, the
following differential equation is obtained 
\begin{eqnarray}
    a ~G'(a) ~\theta~ P_1(x_1) 
        + x_1~ P'(x_1) = 0~.
\end{eqnarray}
The solution is given by
\begin{eqnarray}
    P_1(x_1) = C_1 ~{x_1}^{-a ~G'(a) ~\theta}~.
    \label{HandM}
\end{eqnarray}
Here we consider two cumulative probability distributions
$P_1(> x_1) \propto {x_1}^{-\mu_1}$ and $P_2(> x_2) \propto {x_2}^{-\mu_2}$,
which lead
\begin{eqnarray}
    P_1(x_1) &=& C_1 ~{x_1}^{-\mu_1-1}~,
    \label{Pareto0}\\
    P_2(x_2) &=& C_2 ~{x_2}^{-\mu_2-1}~.
    \label{Pareto2}    
\end{eqnarray}
From Eq.~(\ref{HandM}), (\ref{Pareto0}) and (\ref{Pareto2}),
the relation between $\mu_1$, $\mu_2$ and $\theta$ is expressed as
\begin{eqnarray}
    \frac{\mu_1+1}{\mu_2+1} = \theta~.
    \label{Ratio2}  
\end{eqnarray}

\begin{figure}[htb]
 \centerline{\epsfxsize=0.8\textwidth\epsfbox{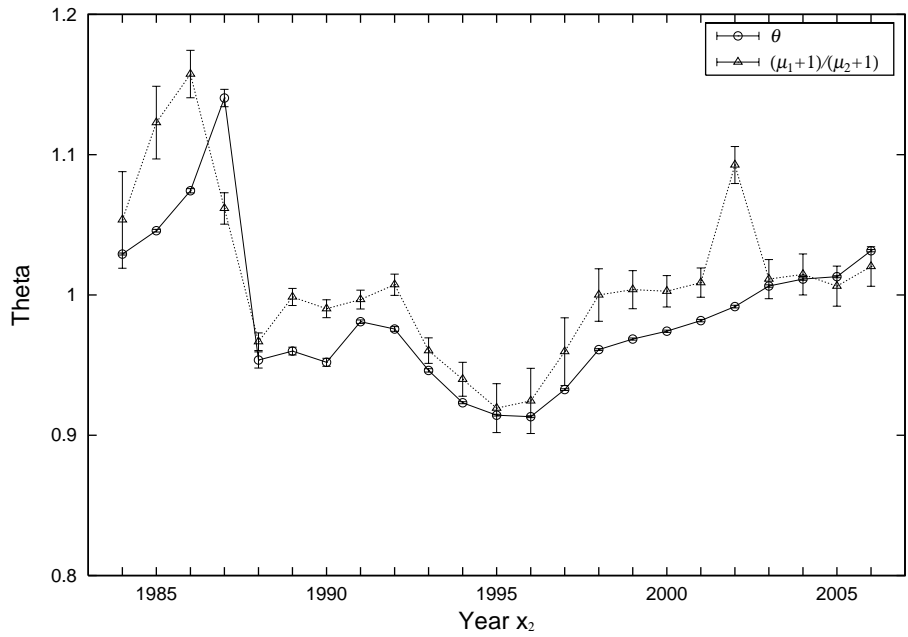}}
 \caption{Annual change of $\theta$ and $(\mu_1+1)/(\mu_2+1)$
 in the year $(x_1, x_2) = (1983, 1984)$ -- $(2005, 2006)$.
}
 \label{Ratio}
\end{figure}

This is an equation between detailed quasi-balance and Pareto's law
in the non-equilibrium dynamical system.
We confirm that the empirical data satisfy this correlation 
in Fig.~\ref{Ratio}.

\section{Conclusion}
\label{sec-Conclusion}
\indspace
In this study,
we have investigated the dynamical behavior
of non-equilibrium system in the large scale region
by employing data on the assessed value of land
in 1983 -- 2006 Japan.
We have identified the detailed quasi-balance (\ref{Detailed quasi-balance})
in the database,
and have derived Pareto's law with varying Pareto index
by assuming Gibrat's law (\ref{Gibrat}).
As a result, 
we have obtained 
a relation between the change of Pareto index $\mu$ and the parameter $\theta$ in
the detailed quasi-balance.
The relation (\ref{Ratio2}) has been confirmed in the empirical data nicely.

\begin{figure}[htb]
 \centerline{\epsfxsize=0.8\textwidth\epsfbox{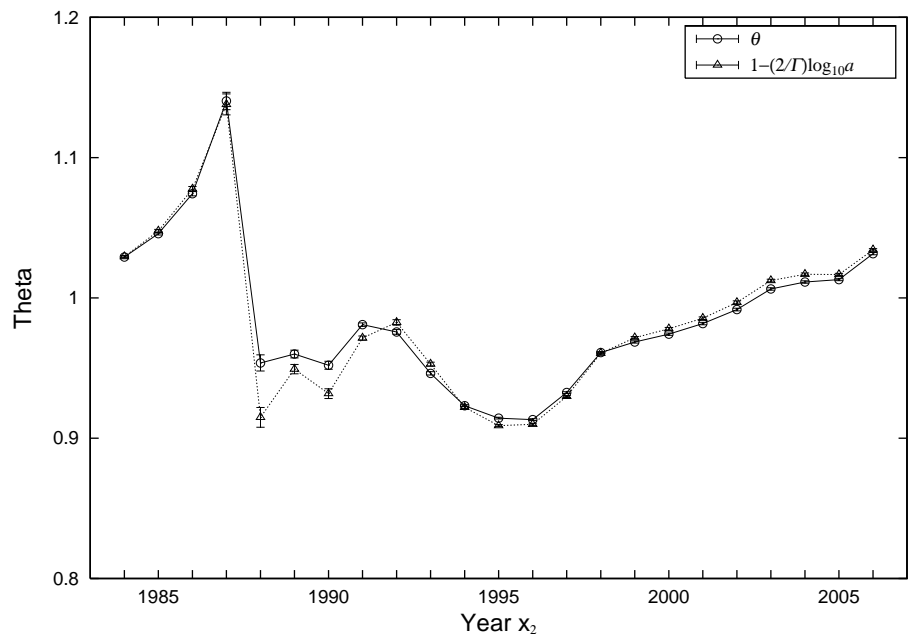}}
 \caption{Annual change of $\theta$ and $1-2/\Gamma \log_{10} a$
 in the year $(x_1, x_2) = (1983, 1984)$ -- $(2005, 2006)$.
 }
 \label{Ratio0}
\end{figure}

What does the other parameter $a$ mean?
Because we demand detailed quasi-balance in the system, 
the area above the regression line (\ref{Line}) equals the area below it.
The two parameters $\theta$ and $a$ are, therefore, related to each other.
The relation is expressed as
\begin{eqnarray}
    \theta = 1 - \frac{2}{\Gamma} \log_{10} a~.
    \label{Gamma}  
\end{eqnarray}
Here $10^{\Gamma}$ is sufficient large number compared with 
the upper bound ($10^7$) where $\theta$ and $a$ are estimated. 
This is the reason why the two parameters $\theta$ and $a$ vary in opposite direction
in Fig.~\ref{DqB}. 
The relation (\ref{Gamma}) is confirmed directly in Fig.~\ref{Ratio0} where
we set $\Gamma$ to be 10.
Consequently, the detailed quasi-balance has one parameter
in principle.

We should comment on several separations between $\theta$ and $(\mu_1 + 1)/(\mu_2 + 1)$ 
in Fig.~\ref{Ratio}.
An abrupt jump of Pareto index between 1984 and 1986 (2001 and 2002)
is observed in Fig.~\ref{VaryingParetoIndex}.
This means that the system changes vigorously
in this period, where
the symmetry is not represented as the regression line (\ref{Line}).
Nevertheless, the dynamical equation (\ref{Ratio2}) is
valid in almost all the other quasistatic periods.

For the next step,
we should investigate the dynamical behavior in the middle scale region.
In Ref.~\cite{Yakovenko},
it is reported that Pareto index $\mu$ changes annually
whereas the distributions in the middle region are stationary in time
by analyzing income data for 1983 -- 2001 USA.
This phenomenon is explained by the Fokker-Planck equation~\cite{F-P}
under two assumptions with respect to the change of income.  
The middle scale distributions of land prices, however,
are not stationary.
The distributions do not collapse onto a single curve
by the normalization of the average land price.
This difference is thought to be caused by the difference between
the trend of increasing (decreasing) income and 
the trend of increasing (decreasing) land price.
In order to study the dynamical behavior in the middle scale region,
we must identify each peculiar feature of middle scale distributions 
in the database~\cite{Ishikawa3}.

\section*{Acknowledgments}
\indent

The author is grateful to 
Professor~V.M.~Yakovenko for useful discussions
about his work,
and to Professor~T.~Kaizoji for useful comments.





\end{document}